# Including Non-Autistic Peers in Games Designed for Autistic Socialization

Yiqi Xiao[1][0009-0009-8725-9106]

[1] University of Illinois Urbana-Champaign, Champaign, IL 61802, USA
`yiqix3@illinois.edu`

**Abstract.** Through a review of current game practices, the author highlights concerns regarding the safety of public social games and the singular medical approach to serious game design for autism. The paper identifies a disconnect between the needs of autistic children and the existing solutions. To fill this gap, a neurodiversity approach to serious game design is proposed. This approach aims to address the social needs of autistic children, enabling them to interact with their neurotypical peers directly, confidently, and safely.

**Keywords:** Autistic Children, Neurodiversity, Social Skills, Educational Game.

## 1 Introduction

### 1.1 Digital Spaces as Alternative Social Environments

Digital games present an important opportunity to support the growing number of autistic children in connecting them to their peers. According to the Centers for Disease Control and Prevention (CDC), the prevalence of autism among 8-year-old children in the United States is 1 in 36 in 2023, a significant increase from 1 in 150 in 2000 [1]. While youth generally enjoy the digital world, Paulus et al. [2] found that autistic boys aged four to seventeen play 85 minutes of video games per day, significantly more than their non-autistic peers. Simpson et al. [3] and Eversole et al. [4] highlighted that digital gaming is one of the most preferred at-home activities for autistic children from both the children's and parents' perspectives.

Ghanouni et al. [5] noted that autistic children often struggle with using language and recognizing social cues, such as picking up on jokes, maintaining eye contact during conversations, or understanding others' feelings through facial expressions or tone of voice [6]. This leads to reduced or atypical interactions between autistic children and their neurotypical peers, which may result in low self-esteem and a high risk of social isolation [7] and bullying at school [8]. To distance themselves from real-world challenges, autistic children often seek respite from the virtual world. Ringland [9] suggests that digital spaces provide an alternative where they can connect with peers, gain confidence, and forge a sense of community. In these digital environments, they can experience interactions and social connections that may be limited in the physical world due to being perceived as "disabled" and socially isolated.

## 1.2    Safety Concerns about Public Social Games

It is important to note that the existence of cyberbullying—the use of technology to harass, threaten, embarrass, or target another person [10]—makes autistic children's experiences in the virtual world remain concerning. According to the Annapolis Police Department, nearly 42% of kids have been bullied online and almost one in four have had it happen more than once [11]. Evidence shows that autistic children face significantly more online safety risks compared to their peers [12]. Digital environments created for public entertainment are not inherently designed for neurodiverse communication and cannot safeguard their wellbeing in social setting with non-autistic individuals.

# 2    Limitations in the Existing Serious Game Practice

## 2.1    Serious Games for Autism Exclude Non-Autistic Children

Learning that public social games may not safely facilitate social interactions between autistic children and their non-autistic peers, it is natural to turn our attention to the field of serious games. While there are indeed many creative practices aimed at addressing the social challenges of autistic children [13, 14, 15], these games offer a singular approach to addressing the social challenges faced by autistic children—attempting to fix them. In the existing games, the activities typically involve fostering language speaking [16], recognizing facial expressions [18], maintaining eye contact [17], and learning appropriate actions or language to use in daily scenarios [19]. All the activities are designed based on social norms, training autistic children to behave more likely to their neurotypical peers.

Take TeachTown [19], a computer-assisted program that teaches young autistic children social and academic skills through games for example. The program transforms numerous social scenarios into engaging animations, helping players understand social norms and expectations in situations such as participating in group discussions or initiating conversations with peers. Following the animations, an interactive activity involves students completing a test on the social rules they have just learned. Games like TeachTown do a good job of making the process of behavior change more bearable and joyful for children by engaging in visuals and interactions. However, at the core of those games, they convey a message to the children: "Your condition is an impairment."

Currently, the vast majority of these social games are designed only for autistic children to use. Among the approximately 60 serious games for autism that are reviewed in three review papers [13, 14, 15], there are about 44 games that aim to address the social challenges, and 0 of them intended to include both autistic children and their non-autistic peers to play together. While such games reduce safety concerns arising from autistic children's interactions with their neurotypical peers, they also potentially exacerbate social isolation and reinforce societal prejudices against autistic children.

### 2.2    The Medical Approach to Serious Game Design

The above approach aligns with the long-standing medical perspective on "treating" autistic children. DSM-5-TR, an international guideline for diagnosing mental health disorders, characterizes autism, or ASD, as a neurodevelopmental disorder [20]. Interventions for autism have traditionally been based on this medical model, aiming to "order the disorder" by providing various training and therapies to improve social and language skills so that they can meet public social standards [21, 22]. Though this approach has been proven effective in improving skills in many areas, it has been widely debated in recent years. Extensive evidence suggests that it can lead to feelings of self-repression, self-deprecation [23], lack of autonomy [24], and thwarted belongingness [25] when autistic individuals feel that they cannot meet expectations or are forced to behave in ways that do not align with their natural tendencies.

Influenced by the medical model, serious game designers typically associate the autism community with a need for extra education. Games designed based on this singular perspective may show effectiveness in measurable behaviors—children learn skills and knowledge after playing the game —but could potentially cause harm on a less observable psychological level. Unlike the extensive research on the harm caused by therapies, there is currently limited research on the negative impacts of therapy-based games on children. This may result from the fact that, among all treatments implying "your condition is an impairment," games are already the most enjoyable parts.

In summary, neither public social games that allow free interactions nor serious games designed solely for autistic children to practice social skills can provide a virtual environment for social interactions between autistic and non-autistic children while ensuring the well-being of autistic children.

## 3    A  New Approach to Social Game Design

### 3.1    The Neurodiversity Approach to Autism

A major misconception is that only autistic children need to connect to their peers. Jones [26] examined the contributors to friendship satisfaction and found that the most important factors are self-disclosure, trust, and companionship enjoyment. Another study, involving 2,555 autism experts from 43 countries in 10 disciplines, identified individuals with autism as having a strong sense of morality, trustworthiness, loyalty, and kindness, indicating that autistic individuals have much to contribute to friendships [27]. It would be a loss if neurotypical children missed the opportunity to have such a reliable friend due to a lack of opportunity to connect and get to know them.

While the strengths of autistic children are not often highlighted and emphasized in the medical model, the neurodiversity approach, a relatively new approach, highlights the importance of recognizing the value of neurodivergent individuals in society. Instead of viewing autism as a disorder, disability, or impairment, this approach sees it as a natural variation of neurotypes [28]. People of different neurotypes need to respect and accept each other's unique mental conditions to support one another. The neurodiversity approach, as emphasized by Singer [29] and Chapman [30] is not a fixed, singular framework but rather a call for ongoing discussion and evolution of the idea. Compared to the medical approach, the discussion of the neurodiversity approach is limited, especially in game design. It is crucial to raise awareness of neurodiversity among practitioners who can make an impact in the autistic community by addressing why we need to change, and demonstrating how we could make a change. The goal of the rest of this paper is to establish such neurodiversity principles to serious game design that address the social needs of autistic children while ensure their wellbeing.

### 3.2     Three Fundamental Principles

This paper proposes three fundamental principles that can guide the design of games supporting neurodiverse socialization: inclusiveness, affirmation, and safety.

- Inclusiveness [39] requires involving all parties in the expected social interactions within the game, aligning with the idea of equity emphasized in the neurodiversity approach. This means that the game's audience should not be limited to only autistic children.
- Affirmation [40], or "neuro-affirming," ensures that the activities in the game allow players to feel comfortable and confident in expressing their neuro-identity. This usually means that game tasks are not designed to highlight the challenges of autistic children.
- Safety here refers only to avoiding the risks brought by interactions between autistic children and their neurotypical peers due to differences in social behavior caused by different neurotypes. This requires both a well-designed internal game mechanisms and an enhanced external supervision.

### 3.3     Case Study I: Inclusive, Safe, but Not Affirming

Sturm et al. [31] documented an innovative participatory design process of a hybrid game aimed at promoting complex emotion recognition and collaboration between autistic individuals and their peers. Players first assemble pieces of facial expressions in a digital puzzle independently and then communicate in person to agree on the appropriate emotion on the face of the body. This project design recognizes the necessity of connecting autistic children with non-autistic children and successfully creates a collaborative activity. It is "inclusive". Although the concept and measurement of safety are not mentioned in the article, the process can be considered "safe" because the experiment was conducted with a very small number of participants and under the supervision of researchers. However, the goal of the activity, "agree on the appropriate emotion," continues to encourage autistic children to modify their social behavior to align with societal expectations. Therefore, although not discussed in the paper, this process might bring stress and discomfort and still follows the medical approach. Besides making the games "inclusive," it is also crucial to ensure that the game activities are "affirming," where autistic children can feel comfortable and confident being themselves.

### 3.4    Case Study II: Affirming, Safe, but Not Inclusive

Battocchi et al. [32] presented the development and evaluation of a collaborative puzzle game for autistic children that is considered "affirming" because it focuses on connecting children through a collaborative experience rather than merely teaching social skills or language. The game employs a tabletop interface and an interaction rule that requires both players to simultaneously touch and drag puzzle pieces. It was tested separately on groups of autistic and non-autistic children, with results demonstrating its effectiveness in promoting social interactions, evidenced by a noticeable increase in simultaneous movements. Therefore, although the game primarily fosters collaboration, it is within the autistic community and thus not considered "inclusive." The process is deemed "safe" because it does not involve interactions between autistic and non-autistic children. Although it does not align with the principle of inclusiveness, it offers valuable insights into designing games for neurodiverse connections. Collaborative activities, such as accomplishing tasks together or creating something jointly, can be highly beneficial as they create more opportunities for communication. Interactions that do not rely on language, such as using body movements or images, reduce communication barriers and allow children with different language abilities to participate confidently.

### 3.5    Case Study III: Affirming, Safe, and Partially Inclusive

Possibly due to the inherent difficulty in creating a game that is inclusive, affirming, and safe, no perfect example embodies this idea that has been found. However, Autcraft [38] offers valuable insights into how developers can make a game affirming while maintaining a degree of inclusiveness for safety purposes.

Minecraft is a popular game among adults and children worldwide, including those with autism. It provides an open world with a lot of freedom for players to choose their activities, such as collaboratively building a city, playing survival games, or simply exploring while chatting [33]. This setting is "affirming" as it offers so much freedom which gives the players much control of what they do in the game. However, this flexibility can lead to rampant cyberbullying. People often enter young players' worlds and destroy their structures, sometimes causing distress by trapping them with lava. Numerous videos of such incidents exist on YouTube [34, 35]. Stuart Duncan [36], an autistic Minecraft player, noted "kids with autism would go onto public servers to play Minecraft and they would just be bullied" in the conversation with the Bridge.

Although Minecraft is not specifically designed for autistic individuals or other neurodiverse people and is thus not inherently a safe environment for them, Duncan saw an opportunity. He explained, "I've often heard it said that people with autism gravitate towards it because all the rules of the world are set and predictable." Building on Minecraft, he developed a server called Autcraft that offers limited access to autistic individuals, their families and friends. In Autcraft, autistic young people can freely and confidently interact with others who share similar experiences or understand the autistic community. To ensure this safe environment, Duncan also appointed parents as admins, who can mute people, temporarily jail them in the game, or ban players for serious rule violations.

As of December 2023, Autcraft has over 17,000 players [37], demonstrating the effectiveness of these measures. In Autcraft, Stuart limited access to non-autistic individuals who are family or friends of autistic individuals to ensure safe interactions and also used strict supervision over the server. However, there are not the only ways to handle it. We could also implement game mechanisms that restrict certain forms of social behavior among players, such as the function of public or private speech, while directing their attention to other well-designed forms of interaction. The example of Autcraft is a call for game designers to pay attention to the long-ignored but significant needs of the autistic community for safe and affirming social interactions with their peers.

## 4      Discussion

Although this paper has been emphasizing the limitations and potential harm brought by serious games developed to teach autistic children social skills, it is not intended to deny the advancements and improvements that gamification practices have achieved in medical treatment, nor their effectiveness in providing practical solutions for autistic children and their parents. Instead of criticizing any specific game, this paper challenges the current singular approach of serious games in addressing the complex life challenges of autistic children and seeks to enrich existing knowledge and practices.

This preliminary framework has two major limitations. Firstly, due to a scarcity of relevant work, there is a lack of empirical data or pilot studies to support its claims about the effectiveness of neurodiversity principles in game design. Secondly, it is challenging to implement all three principles in one game. In highly inclusive environments, the level of social safety may decrease, which can cause children to feel uncomfortable even if the activities within the game are designed to be affirming. Although previous sections have provided some actionable insights, more methods need to be discussed to better balance these three principles.

## 5      Conclusion

By reviewing current practices in the digital game area and referring to extensive research on autistic children, this paper highlights a disconnect between autistic children's need to safely and confidently socialize with their peers in the digital world and the existing solutions. To fill this gap, the author advises serious game designers to shift their focus from "teaching autistic children social skills" to "facilitating neurodiverse social interactions" with three main principles: inclusiveness, affirmation, and safety. This preliminary work intends to spark ongoing discussions in further developing these guidelines and enrich the existing knowledge and practices. The neurodiversity approach in serious game design is like an open door; a path will only be forged when more people pass through it.

**Acknowledgments.** The author would like to thank Boyang Wu and Katryna Starks for their valuable assistance in refining the language and improving the clarity of this paper.